\documentclass{desyproc}
\usepackage{axodraw4j}
\usepackage{subfigure}

\begin{document}
%------------------------------------
\title{\vspace{-3cm}
\hfill{\small{IPPP/10/91; DCPT/10/182}}\\[2cm]
%\hfill{\small{DCPT/08/114}}\\[1cm]
Low energy laboratory searches for WISPs}

%for single authors the superscripts are optional
\author{{\slshape Joerg Jaeckel}\\[1ex]
Institute for Particle Physics Phenomenology, Durham University, Durham DH1 3LE, United Kingdom\\[-0.5cm]}

% if the proceedings are available online (e.g. at Indico)
% please enter the contribution ID or file_name below for the DOI
%\contribID{32}
%\contribID{Jaeckel_Joerg}

% TO THE CONFERENCE EDITORS: 
% please update the following information      
% before sending the template to the authors
% \confID{800}  % if the conference is on Indico uncomment this line
\desyproc{DESY-PROC-2010-03}
\acronym{Patras 2010} % if you want the Acronym in the page footer uncomment this line
%\doi  % if there is an online version we will register DOIs

\maketitle

\begin{abstract}
We present an overview over the current status of laboratory experiments searching for (very) weakly coupled slim particles (WISPs).
These experiments at the high precision frontier explore new physics beyond the standard model in a complementary way 
to high energy accelerators. 
The multitude of active and planned experiments shows a lively field and promises interesting new data in the near future.
\end{abstract}

\section{Introduction}\label{Introduction}
Exploring new physics in its very meaning requires that we go beyond the boundaries of what is known. 
For example the Large Hadron Collider (LHC) will push the high energy frontier into the multi-TeV range, exploring untested regions with the promise
of exciting discoveries.
At the same time there is actually a wide range of experiments pushing in a different, complementary direction by going to high precision.
In this note we will briefly review a range of these high precision experiments, in particular those looking for new light, (very) weakly interacting particles,
called weakly interacting slim particles (WISPs). 

When searching for new particles the natural question to ask is: why haven't we already seen them?
The high energy and the high precision frontier follow from the two different possible answers to this question. One possible answer is that the particles are very massive. Then we simply do not have enough energy to produce them. The solution are high energy experiments.
Alternatively the particles interact only very weakly with ordinary matter, then we have a hard time producing and detecting them, even if they are very light. In this case high precision searches are more promising. 

For the high energy frontier we have a lot of circumstantial evidence that points us towards the existence of new physics at the TeV scale explored at the LHC. But do we also expect new physics at low energies?
The answer to this question is yes. Indeed we already have observed puzzling phenomena connected to very low energy scales:
\begin{itemize}
\item{Neutrinos have masses (or more precisely mass splittings $\Delta m^2$) of the order of meV.}
\item{The energy density of dark energy (at best very weakly coupled to ordinary matter) is of the order of $(\rm meV)^4$.}
\item{The total energy density of the universe (of which dark energy is about 70 \%) is of the order of $(\rm meV)^{4}$, too\footnote{The surprising thing about this is that the matter density is of the same order of magnitude as the dark energy density at the present time. This is the so-called coincidence problem.}.}
\end{itemize}
Moreover, there is quite often an interesting connection between underlying physics happening at very high energy scales, to low masses
and very weak couplings (see, e.g.,~\cite{Jaeckel:2010cz}). A prime example for this connection are (pseudo-)Goldstone bosons arising from spontaneous symmetry breaking
of a global symmetry. If $M$ is the scale at which the symmetry is broken their (effective) couplings are typically suppressed by powers of $1/M$.
For true Goldstone bosons the masses vanish and for pseudo-Goldstone bosons the mass is again suppressed by the symmetry breaking scale, $\sim \Lambda^2/M$. The classic example of such a pseudo-Goldstone particle, originating from a spontaneously 
broken U(1) symmetry proposed to solve the strong CP problem, is the 
axion~\cite{Peccei:1977hh,Weinberg:1977ma,Wilczek:1977pj,Kim:1979if,Dine:1981rt,Shifman:1979if,Zhitnitsky:1980tq}. The axion is still one of the prime
WISP candidates and finding it is a strong motivation for many of the experiments discussed below. 

With this motivation in mind let us now return to the main topic of this note. How can we search for WISPs with their feeble interactions?
In the following sections we will present an overview of the techniques and existing experiments to search for WISPs. 
For brevity we will focus on a particular class of WISPs interacting with photons (for 5th forces experiments 
see~\cite{Hoedl:Zuerich,Adelberger:2009zz}).  

\section{Light shining through walls}
\begin{figure}[t!]
\begin{center}
  \scalebox{0.5}[0.5]{
  \begin{picture}(386,130) (159,-207)
    \SetWidth{1.0}
    \SetColor{Black}
    \GBox(336,-206)(368,-78){0.882}
    \Line[dash,dashsize=8.4,arrow,arrowpos=0.5,arrowlength=5,arrowwidth=2,arrowinset=0.2](272,-142)(432,-142)
    \Photon(160,-142)(272,-142){7.5}{6}
    \Photon(432,-142)(544,-142){7.5}{6}
    \Vertex(432,-142){8}
    \Vertex(432,-142){8}
    \Vertex(272,-142){8}
    \SetOffset(0,-10)
    \Text(224,-110)[c]{\Large{\Black{$\gamma$}}}
    \Text(480,-110)[c]{\Large{\Black{$\gamma$}}}
    \Text(304,-110)[c]{\Large{\Black{$X$}}}
    \Text(400,-110)[c]{\Large{\Black{$X$}}}
  \end{picture}
  }
  \end{center}
\caption{
Schematic of a ``light-shining-through a wall'' experiment.
An incoming photon $\gamma$ is converted into a new particle $X$
which interacts only very weakly with the opaque wall. It passes
through the wall and is subsequently reconverted into an ordinary
photon which can be detected.
}\label{fig:lsw}
\end{figure}
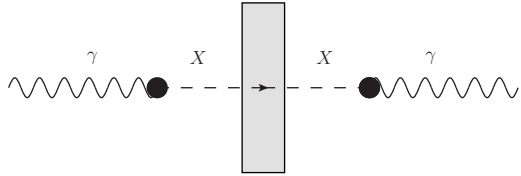

Light shining through walls (LSW)~\cite{Okun:1982xi,Anselm:1986gz,VanBibber:1987rq} (for a review see~\cite{Jaeckel:2010ni,Redondo:2010-review}) is one of the main techniques to search for WISPs in the laboratory.
The basic process (shown in Fig.~\ref{fig:lsw}) really does what the name suggests, it allows photons to traverse
a completely opaque wall. The photon basically avoids the wall by converting into particle(s) which interact only very weakly with ordinary matter.

%%%%%%%%%%%%%%%%%%%%%%%%%%%%%%%%%%%%%%%%%%%%%%%%%%%%%%%%
%%%%%%%%%%%%%%%%%%%%%%%%%%%%%%%%%%%%%%%%%%%%%%%%%%%%%%%%

The power of this approach becomes immediately evident when considering the following numbers.
A laser with a wavelength of 1064~nm and a power of only 20~W produces of the order of $10^{20}$ photons per second.
On the other side, detecting 1 photon per second is quite doable. Therefore even such a simple experiment
can test probabilities for the process $\gamma\to X\to \gamma$ of the order of $P_{\gamma\to X\to\gamma}\sim 10^{-20}$. 
This makes it quite clear that we are able to test very weak couplings. The price to pay is that the energy to produce a particle $X$ is typically
of the order of the energy of the incoming photon and therefore $\sim$~eV. In other words we can only produce very light particles with masses
$\lesssim$~eV.

So what can the particle $X$ be? And how do we achieve the conversion? Let us consider here only the two simplest possibilities.
A (pseudo-)scalar $\phi$ could couple to two photons via a term $\sim g/4\phi (F^{\mu\nu})^2$ ($\sim g/4\phi F^{\mu\nu}\tilde{F}_{\mu\nu}$).
This allows the conversion of photons into these so-called axion-like particles (ALPs) in the presence of a background magnetic field.
Alternatively we could have spin-1 gauge bosons just like the ordinary photon but hidden by the fact that all ordinary matter particles carry no charge under this ``hidden photon''. The interaction then takes place via a kinetic mixing term~\cite{Okun:1982xi,Holdom:1985ag} in the Lagrangian $\sim \chi F^{\mu\nu}X_{\mu\nu}$ where $X^{\mu\nu}$ is the field strength corresponding to this hidden photon and $\chi$ gives the mixing 
angle\footnote{Such a mixing arises quite naturally in both field~\cite{Holdom:1985ag}
and string theory~\cite{Dienes:1996zr,Abel:2003ue,Abel:2006qt,Abel:2008ai,Goodsell:2009xc,Bullimore:2010aj}.}. If the hidden photons have a small mass (this can be generated either via a Higgs or via a Stueckelberg mechanism) then we have photon -- hidden photon oscillation which are completely analogous to neutrino oscillations.

\begin{figure}[t!]
\centerline{
\subfigure[]{
\begin{picture}(200,130)
\includegraphics[width=.48\textwidth]{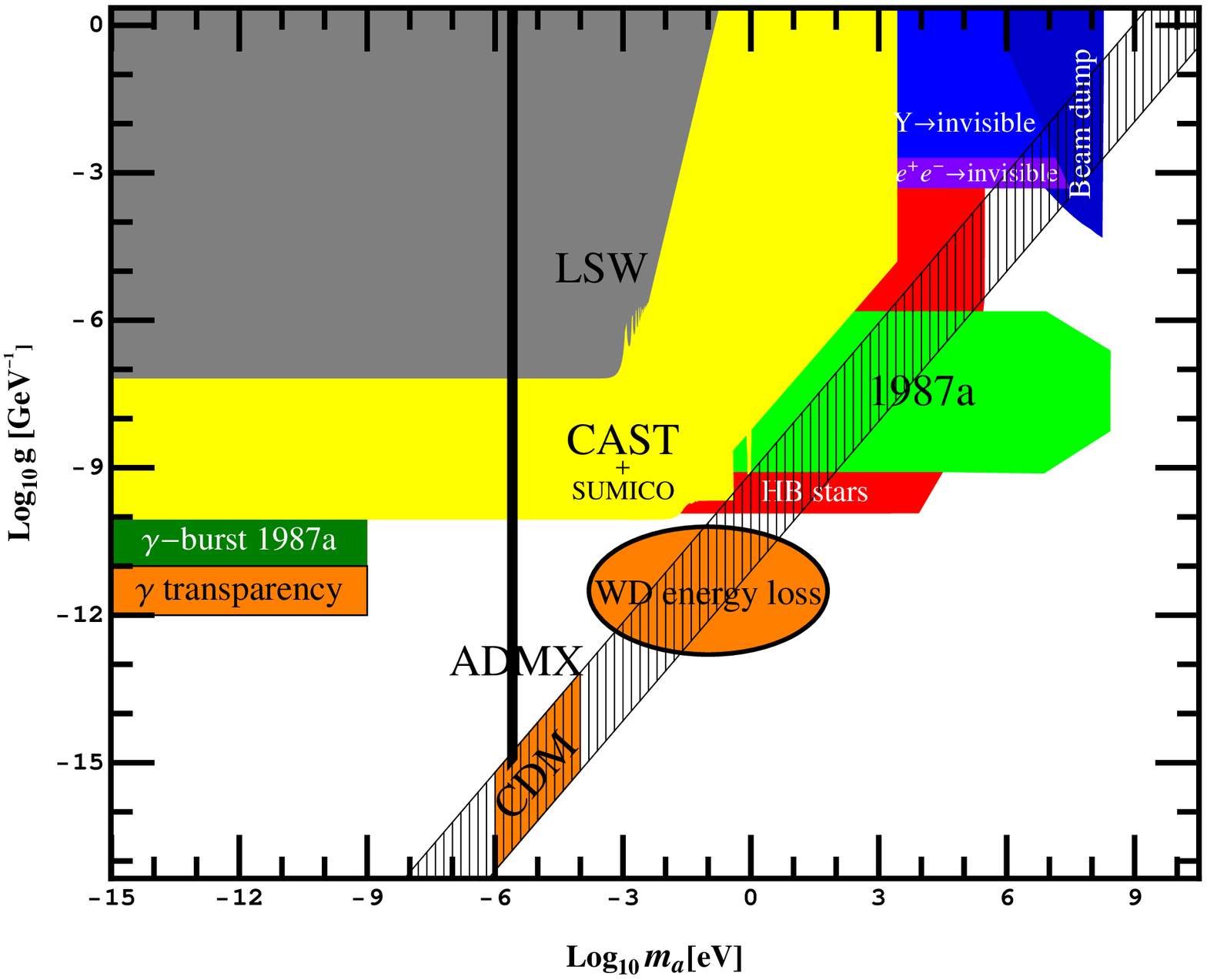}
\end{picture}
\label{alpbounds}}
\hspace*{0.2cm}
\subfigure[]{\begin{picture}(200,130)(0,-30)
\includegraphics[width=0.46\textwidth]{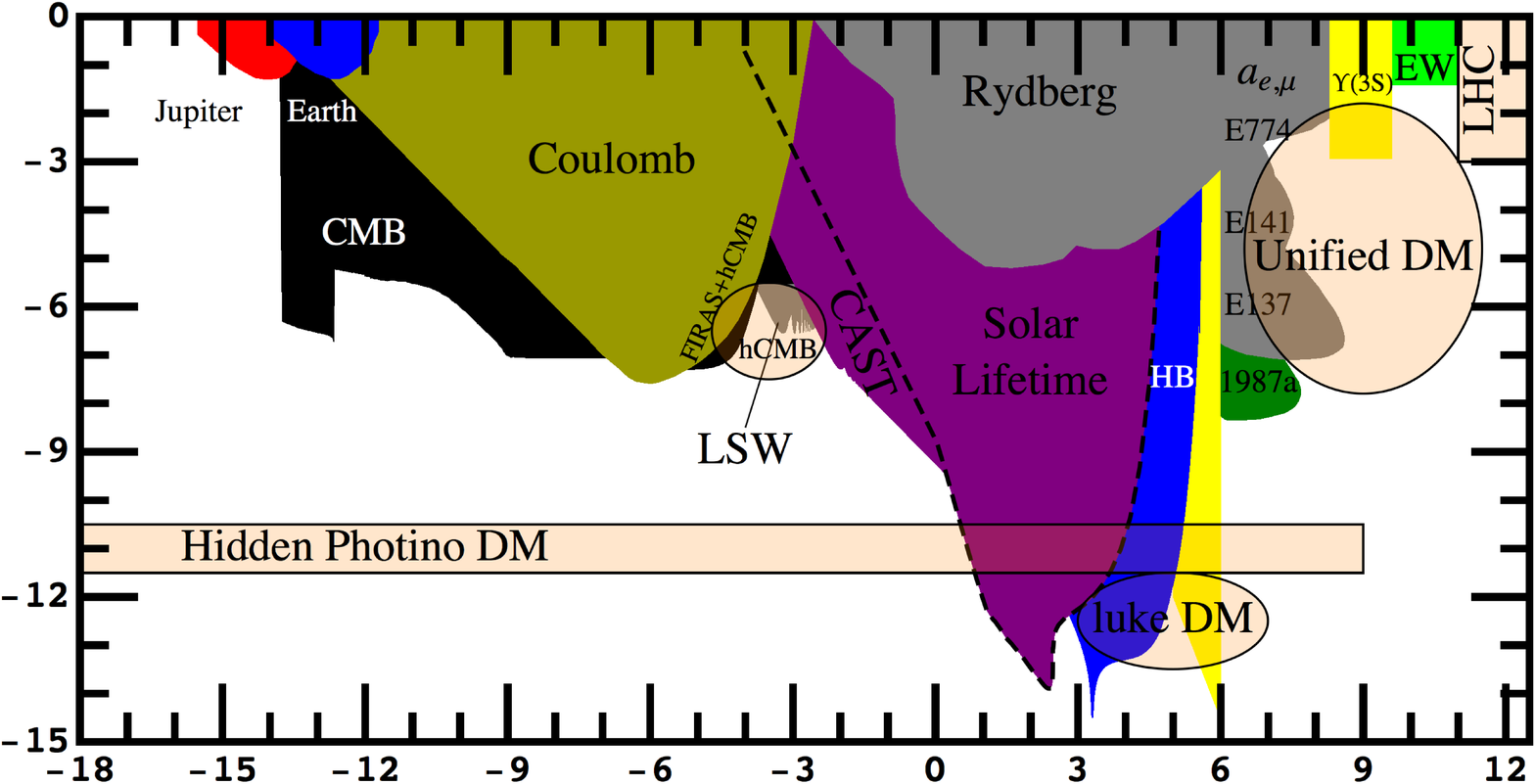}
\rText(-190,51.5)[c][l]{\scalebox{0.6}[0.6]{${\rm Log}_{10}\chi$}} 
\Text(-90,-5)[c]{\scalebox{0.6}[0.6]{${\rm Log}_{10}m_{\gamma^{\prime}}[{\rm eV}]$}} 
\end{picture}\label{hpbounds}}
}
\caption{Summary of constraints
for axion-like particles (left panel) and hidden photons (right panel). Both compilations taken from~\cite{Jaeckel:2010ni} where also details can be found. 
The mass region, where the axion is a natural dark matter candidate is marked in orange and labeled  ``CDM". 
Also other areas which are especially interesting are marked in orange (transparent in (b)). 
}\label{bounds}
\end{figure}

Experimentally the last few years have brought enormous progress. After the pioneering BFRT experiment in the early 1990s the last three
years have brought a flurry of activity and significant improvements with a large number of experiments 
ALPS~\cite{Ehret:2009sq,Ehret:2010mh}, BMV~\cite{Fouche:2008jk}, GammeV~\cite{Chou:2007zzc}, LIPSS~\cite{Afanasev:2008fv,Afanasev:2008jt} and OSQAR~\cite{Pugnat:2007nu}
taking data, publishing results and improving their setup on the time-scale of month. Since unfortunately no light shining through walls was found the 
results are in the form of upper bounds which are summarized in Fig.~\ref{bounds} for ALPs and hidden photons.
We can see that these experiments are indeed sensitive to tiny couplings!

From Fig.~\ref{bounds} the challenges for the future are clear. In particular for ALPs the helioscope bounds (discussed below) are still significantly better. So one of the goals will be to improve the sensitivity of LSW experiments beyond this. Moreover, it would also be desirable
to extend the reach towards larger masses in order to come closer to the predicted region 
for the QCD axion (the hatched diagonal band in Fig.~\ref{alpbounds}).
The first goal might be facilitated by using more magnets~\cite{Ringwald:2003nsa} and by employing optical cavities in both the production as well as in the regeneration region~\cite{Hoogeveen:1990vq,Hoogeveen:1992uk,Sikivie:2007qm,Jaeckel:2007ch}.
This ``resonant regeneration'' enhances the number of photons reaching the detector by a factor proportional to the number of
passes the light does in each cavity $~N_{\rm pass,\, prod}N_{\rm pass,\, reg}$ (the ALPS experiment already pioneered the use of an optical cavity in the production region~\cite{Ehret:2009sq,Ehret:2010mh}). Moreover, improvements in detection sensitivity will further help.
Higher masses can be reached using suitable configurations of several magnets and other tricks~\cite{VanBibber:1987rq,Arias:2010bh,Arias:2010bha}. However, to reach the axion line further improvements may be necessary.

For hidden photons the situation is significantly better in that already the current experiments provide the best bounds for masses in the interesting region of a meV. The improvement discussed above will significantly increase the probed area of parameter space providing huge discovery potential for new physics.

The idea of resonant regeneration might work even better with ``light'' in the microwave region. Here the light can be reflected up to $10^{11}$ times inside the cavity, promising interesting sensitivities~\cite{Hoogeveen:1992uk,Jaeckel:2007ch}. Again several groups 
are planning and or building such an experiment~\cite{Slocum:2010,Caspers:2010,Williams:2010} and two test setups have already yielded first data~\cite{Povey:2010hs,Wagner:2010mi}.

Modified LSW setups, so called afterglow experiments can be used to search for chameleons (a special type of scalar ALP connected to dark energy).
Chameleons are different from ordinary ALPs in that dense regions of ordinary matter, like mirrors or a wall also represent barriers for the chameleon.
Therefore, chameleons produced in a vacuum tube closed with transparent windows on both sides, remain trapped in it. 
The experiment then works as follows. First a laser is shone through the tube, effectively ``charging'' it with chameleons. Then the laser is switched off.
The trapped chameleons then slowly reconvert into ordinary photons which can leave the tube through the windows and be detected.
An experiment searching for this ``afterglow''~\cite{Ahlers:2007st,Gies:2007su} has been performed by 
the GammeV collaboration producing new bounds on chameleons~\cite{Chou:2008gr,Steffen:2010ze}. Moreover, the ADMX collaboration did a test run in the microwave regime~\cite{Rybka:2010ah}.

\section{Helioscopes}
Helioscopes~\cite{Sikivie:1983ip} follow the same principle as LSW experiments. 
The difference is that on the production side the laser photons are replaced by the photons in the core of the sun.
For ALPs the conversion of these photons is stimulated by the electromagnetic fields of the ions in the plasma (as discussed above
such fields are not necessary for hidden photons; nevertheless the conversion is affected by the plasma due to the modified photon propagation).
On earth one then has a reconversion region (oriented towards the sun) just as in an ordinary LSW experiment. Everything between the solar core and this regeneration region effectively can be considered part of a very thick wall.
 
Due to the enormous number of photons inside the solar core this is a very powerful technique. Indeed, as can be seen from Fig.~\ref{alpbounds}
the CAST experiment is setting the best limits on ALPs for masses below  10~meV~\cite{Andriamonje:2007ew}. This is currently many orders of magnitude better then the LSW experiments, thereby setting the benchmark for the next generation of those experiments.
In addition the SUMICO collaboration has produced interesting bounds for masses around 1~eV~\cite{Inoue:2008zp}.
For hidden photons the situation is very good in that the helioscopes yield complementary bounds to the LSW experiments at somewhat higher masses (cf. Fig.~\ref{hpbounds}). Dedicated hidden photon helioscopes such as the proposed SHIPS may even compete with
LSW experiments in the meV range~\cite{Cadamuro:2010ai}.

Unfortunately, helioscopes are also still somewhat shy of the axion region. Future improvements can include a stronger magnet and a wider aperture.
The former enhances the regeneration probability for ALPs. The latter generally increases the WISP flux that can be used
for reconversion and hence also increases the sensitivity for hidden photons.

At higher masses sensitivity is lost, because the
axion wave and the regenerated photon wave\footnote{In LSW experiments the same happens on the production side.} have slightly different
wavelengths. After some distance newly regenerated photons have opposite phase to those produced at an earlier point and interfere destructively, effectively reducing the number of regenerated photons. This problem can be solved by changing the wavelength of the photons to match those
of the WISP. This can be achieved by filling the regeneration region with a buffer gas with a refractive index such that $(n-1)=-m^2_{\phi}/(2\omega^2)$.
In the X-ray regime (photons in the solar core have X-ray energies) refractive indices smaller than 1 can indeed occur in ordinary gases like helium.
To reach higher masses the CAST and SUMICO collaborations therefore fill their regeneration regions with such a 
gas~\cite{Arik:2008mq,Inoue:2008zp}. Higher masses will require more gas in the beam line\footnote{It should be noted that at higher masses more and more tuning steps are required making this increasingly difficult.}.

\section{Haloscopes - Searching for axion dark matter}
True QCD axions (and perhaps to some degree also general axion-like particles) can form all or a part of dark matter (it is actually
one of the most favorite dark matter candidates~\cite{Baer:2010wm}). 
In the axion mass range (1-100)$\mu$eV production of a sufficient number of suitably cold axions is indeed a natural expectation (orange ``CDM'' region in Fig.~\ref{alpbounds}).
This provides a plentiful source of these particles and therefore another opportunity to find them.

Having the dark matter axions all around us we only need to reconvert them and detect the produced photons. This is the idea
of a haloscope~\cite{Sikivie:1983ip}. In a sense
such an experiment would convert dark matter into electricity! 
The advantage of such an experiment is that we pay the price of the small coupling only for the reconversion.
(The disadvantage is that we rely on the additional assumptions that axions are indeed contributing to dark matter.) 

Since dark matter axions are assumed to have very small velocities their energy is basically determined by the rest mass. Therefore, regenerated photons are in the microwave regime. As in the LSW experiments the regeneration probability can be enhanced by employing a high quality cavity.
For this enhancement to be effective the regenerated photons which have energy $\approx m_{a}$ must be resonant in the cavity. Since the mass of the axion is unknown such an experiment must scan through the masses changing the resonance frequency of the cavity in each step.

An experiment of this type is currently being performed by the ADMX collaboration. The data collected in the last few years yields already
an absolutely impressive sensitivity~\cite{Asztalos:2003px,Asztalos:2009yp}, currently being the only laboratory experiment testing true QCD axions. The price to pay is that the
mass range is somewhat limited. 
However, currently ADMX is implementing improvements in the detector sensitivity which will greatly enhance the scanning speed. This
will then allow to scan about a decade in axion masses in a few years.
The CARRACK collaboration is developing microwave detectors based on Rydberg atoms~\cite{Tada:1999tu} which could improve the (already impressive) sensitivity to detect microwave radiation even further.

\section{Laser polarization experiments}
In all the experiments above, detection was based on reconverting the WISPs (produced in the laboratory or from a source like dark matter)
into photons. In that sense they are direct detection or appearance experiments.
But in principle we could also do a disappearance experiment.
This is the principle with which laser polarization experiments can search for WISPs.

The basic idea is that as polarized light passes through an interaction region, the real or virtual production of WISPs can leave traces
in the polarization of the laser light. This is most easily understood in the case of (pseudo-scalar) ALPs. Here the interaction is
given by $\sim g\phi \mathbf{E}\!\cdot\!\mathbf{B}$. In a background magnetic field $\mathbf{B}$ the photons provide the $\mathbf{E}$ field necessary
for the interaction. We can immediately see that only those photons with an $\mathbf{E}$ field parallel to the magnetic field interact.
When an ALP is produced, the corresponding photon is absorbed. Consequently, we have a selective absorption of photons parallel to the magnetic field. This is a so-called dichroism, and leads to a rotation of the polarization direction. Similarly, virtual production causes a polarization dependent effective refractive index (birefringence) which leads to a phase shift between the two polarizations turning an initially linear polarized laser beam into one with an elliptic polarization.

To allow the different polarization directions to be affected differently we need a preferred direction in space. Typically this is provided by a background magnetic field. Therefore polarization experiments are typically not sensitive to hidden photons which do not interact with this magnetic field.
However, in presence of additional hidden matter this can change~\cite{Gies:2006ca,Ahlers:2007rd}.

Measurements of this type have already been performed by the BFRT~\cite{Cameron:1993mr}, PVLAS~\cite{Zavattini:2007ee}, 
Q\&A~\cite{Chen:2006cd} and BMV~\cite{Bielsa:2009dz} collaborations and OSQAR~\cite{Pugnat:2007nu} is also setting up such an experiment.
The sensitivity of polarization experiments is somewhat limited due to difficult to control experimental backgrounds (at least for the rotation signature the theoretical backgrounds are very low~\cite{Ahlers:2008jt}). Nevertheless, these experiments provide a useful platform to develop optical techniques such as Fabry-Perot cavities.
Moreover, and maybe even most importantly they also have an additional physical goal: They want to test the QED prediction of vacuum magnetic birefringence created by virtual electron positron loops.

\section{Tests of Coulomb's law}
Tests of Coulomb's law can also be a very sensitive probe for WISPs. In particular, hidden photons cause  a small Yukawa contribution 
$\sim\alpha \chi^2 \exp(-m_{\gamma^{\prime}}r)/r$ to the Coulomb potential which allows for a very powerful test of these particles as can be seen from Fig.~\ref{hpbounds}.

The yellow ``Coulomb'' region in Fig.~\ref{hpbounds} originates~\cite{Popov:1999} from an experiment performed at a scale of 10s of cm already 39 years ago~\cite{Williams:1971ms}. Therefore, it seems very plausible that with current technology significant improvements are possible.

Tests of Coulomb's law can also be performed at other length scales. Indeed at large length scales the bounds ``Jupiter'' and ``Earth'' in 
Fig.~\ref{hpbounds} arise from ``Coulomb's law'' tests on the magnetic fields of Jupiter and Earth. 
Similarly one can use atomic spectra to probe Coulomb's law at very small length 
scales, labelled ``Rydberg''~\cite{Popov:1999,Pospelov:2008zw,Karshenboim:2010cg,Jaeckel:2010xx} in Fig.~\ref{hpbounds}. 
As a test of hidden photons these bounds are particularly 
model-independent\footnote{For example, the fixed target experiments discussed below typically assume
that hidden photons with mass greater than the electron threshold dominantly decay into electrons (and perhaps muons). This can be different in models with hidden matter.}~\cite{Karshenboim:2010cg,Jaeckel:2010xx}.

\section{Searches at higher masses}
Over the last few years, a variety of dark matter experiments and astrophysical observations have provided interesting hints
towards hidden photons with somewhat higher masses in the MeV-GeV range\footnote{These masses are still below the current high energy frontier $\sim$TeV and particles in this range must have (moderately) small couplings to the standard model in order to be viable. Accordingly experiments searching for them must be sufficiently precise and we include them in our discussion of the high precision frontier.}
 (labelled ``Unified DM''~\cite{ArkaniHamed:2008qn} in Fig.~\ref{hpbounds}).
Similarly, hidden photons with masses in this range could also provide an explanation~\cite{Pospelov:2008zw} for the deviation of $(g-2)_{\mu}$ from
the standard model prediction~\cite{Hagiwara:2006jt}.
Moreover, masses in this range could arise quite naturally in 
supersymmetric~\cite{Chun:2008by,Cheung:2009qd,Morrissey:2009ur,Baumgart:2009tn,Cui:2009xq} or string theory setups~\cite{Goodsell:2009xc}.

Hidden photons (and typically also ALPs) at masses above the electron threshold, $2m_{e}\approx1\,{\rm MeV}$, can be nicely
probed in fixed target experiments~\cite{Bjorken:2009mm}. 
Here a high current beam of electrons (possibly also protons) is shot onto a target, typically a sheet or even a block of metal.
Hidden photons are then produced, e.g. via hidden photon Bremsstrahlung. Depending on the size of the kinetic mixing the hidden photons can
then  travel a significant distance before they decay into electron-positron pairs (or if sufficiently massive muons) which can be detected.

A wide range (see~\cite{Andreas:2010tp} for an overview) of these experiments is currently in planning, testing or early operation states at various accelerator facilities such as DESY (HIPS~\cite{Mnich:2010}), MAMI~\cite{Denig:2010} and Jefferson Lab (APEX~\cite{Essig:2010xa}, HPS~\cite{Maruyama:2010} and DarkLight~\cite{Fisher:2010}).

Moreover, these ``heavyish WISPs'' can also be searched for in meson experiments~\cite{Pospelov:2008zw,Reece:2009un,Batell:2009jf,Andreas:2010ms}.

\section{Conclusions}
High precision experiments at low energies have great potential to search for new light, but very weakly coupled particles.
Thereby, they can shed light on the so-called ``hidden sector'' and deliver complementary information to high energy collider
experiments.

A multitude of active and near future experiments will explore this high precision frontier in the next years, bringing a wealth of new information
on the nature of fundamental physics and hopefully exciting discoveries.

\section*{Acknowledgements}
The author would like to thank L.~Baudis, M.~Schumann and the local organizing team of the 6th Patras-Axion-WIMP-WISP workshop for creating a wonderful atmosphere. Moreover, he would like to thank J.~Redondo and A.~Ringwald for interesting discussions and useful comments.

% ****************************************************************************
% BIBLIOGRAPHY AREA
% ****************************************************************************

\begin{footnotesize}

\bibliographystyle{h-physrev5}
\bibliography{../masterbib}

\end{footnotesize}

% ****************************************************************************
% END OF BIBLIOGRAPHY AREA
% ****************************************************************************

\end{document}